\documentclass[aps,pra,reprint,superscriptaddress]{revtex4-1}
\usepackage[colorlinks=true,allcolors=blue]{hyperref}
\usepackage{amsmath}
\usepackage{amssymb}
\usepackage{bm}
\usepackage[pdftex]{graphicx}
\usepackage{microtype}
\usepackage{verbatim}
\usepackage{todonotes}
\presetkeys{todonotes}{inline,backgroundcolor=yellow}{}
\usepackage{subcaption}
\newcommand{\bp}[1]{\left( {#1} \right)} % "bracket-parenthetical"
\newcommand{\bs}[1]{\left[ {#1} \right]} % "bracket-square"
\newcommand{\bc}[1]{\left\{ #1 \right\} } % "bracket-curly"

\newcommand{\id}{\mathbb{I}}
\newcommand{\ket}[1]{{| {#1} \rangle}}

\newcommand{\ketbra}[2]{{\left| {#1} \right\rangle \!\!\left\langle {#2} \right|}}

\newcommand{\cg}[6]{
\left\langle 
\begin{array}{@{}cc|c@{}}
    #1 & #3 & #5 \\
    #2 & #4 & #6
\end{array}
\right\rangle
}

\DeclareMathOperator{\Tr}{Tr}
\newcommand{\de}{\mathrm{d}}

\begin{document}
%\linenumbers
\title{Wigner negativity in spin-$j$ systems}
\author{Jack Davis}
\affiliation{Department of Physics and Astronomy, University of Waterloo, Waterloo, Ontario, Canada N2L 3G1}
\affiliation{Institute for Quantum Computing, University of Waterloo, Waterloo, Ontario, Canada N2L 3G1}
\author{Meenu Kumari}
\affiliation{Perimeter Institute for Theoretical Physics, 31 Caroline St N, Waterloo, Ontario, Canada N2L 2Y5}
\author{Robert B. Mann}
\affiliation{Department of Physics and Astronomy, University of Waterloo, Waterloo, Ontario, Canada N2L 3G1}
\affiliation{Institute for Quantum Computing, University of Waterloo, Waterloo, Ontario, Canada N2L 3G1}
\affiliation{Perimeter Institute for Theoretical Physics, 31 Caroline St N, Waterloo, Ontario, Canada N2L 2Y5}
\author{Shohini Ghose}
\affiliation{Department of Physics and Computer Science, Wilfrid Laurier University, Waterloo, Ontario, Canada N2L 3C5}
\affiliation{Institute for Quantum Computing, University of Waterloo, Waterloo, Ontario, Canada N2L 3G1}
\affiliation{Perimeter Institute for Theoretical Physics, 31 Caroline St N, Waterloo, Ontario, Canada N2L 2Y5}

\begin{abstract}
The nonclassicality of simple spin systems as measured by Wigner negativity is studied on a spherical phase space.  Several SU(2)-covariant states with common qubit representations are addressed: spin coherent, spin cat (GHZ/N00N), and Dicke ($\textsf{W}$).  We derive a bound on the Wigner negativity of spin cat states that rapidly approaches the true value as spin increases beyond $j \gtrsim 5$.  We find that spin cat states are not significantly Wigner-negative relative to their Dicke state counterparts of equal dimension.  We also find, in contrast to several entanglement measures, that the most Wigner-negative Dicke basis element is spin-dependent, and not the equatorial state $\ket{j,0}$ (or $\ket{j,\pm 1/2}$ for half-integer spins).  These results underscore the influence that dynamical symmetry has on nonclassicality, and suggest a guiding perspective for finding novel quantum computational applications.
\end{abstract}
\maketitle

\section{Introduction}

The phase space formulation of quantum mechanics allows the representation of states and operators as scalar functions on classical phase space.  This is useful for distinguishing between classical and quantum behaviour as well as for experimental state tomography. Beginning in 1932 by Wigner and others, the phase space formulation is now widely used in quantum optics and quantum information \cite{zachos_quantum_2005}.  The Stratonovich-Weyl correspondence is a unified phase space framework that focuses on a system's dynamical symmetry \cite{stratonovich_distributions_1956}.  It allows us to define families of quasiprobability distributions (QPDs) on many classical phase spaces. These include the plane and the sphere, as generated by Heisenberg-Weyl and SU(2) symmetries respectively \cite{brif_phase-space_1999}.

Due to the quantum nature of states in Hilbert space, any QPD will in general fail to be a true probability distribution in the sense of classical statistical mechanics \cite{cahill_density_1969}. The failure can occur in different ways depending on how the QPD is defined \cite{ferrie_frame_2008}.  This has led to the idea of quantifying the nonclassicality of a quantum state by measuring the difference between a QPD representation of the state and a true probability distribution over the same phase space.  For the case of QPDs on a plane, the most widely used is the Wigner function \cite{leonhardt_measuring_1997}. The failure of such a Wigner function QPD to be a true probability distribution is seen in the presence of negative values.  In this case, it is common to use the total negative volume as a quantification of the failure, and interpret this \textit{Wigner negativity} as a measure of nonclassicality \cite{kenfack_negativity_2004}. In this paper, we explore the nonclassicality of quantum spin states using Wigner negativity.

Under suitable circumstances, Wigner negativity has been linked to other measures of quantumness such as entanglement and contextuality \cite{dahl_entanglement_2006, spekkens_negativity_2008, arkhipov_negativity_2018}. The use of phase space methods in quantum computation has also led to several  connections.  For example, Wigner negativity and quantum contextuality have been shown to be equivalent notions of nonclassicality for multi-qudit systems with odd local dimension (though this fails for single qudits) \cite{delfosse_equivalence_2017}.  Furthermore, in odd dimensions, the stabilizer states have nonnegative Wigner functions \cite{gross_hudsons_2006}. This draws a connection to the Gottesman-Knill theorem, which demonstrates the efficient classical simulation of stabilizer circuits \cite{gottesman_heisenberg_1998}.  This result has been extended to the continuous variable setting where it was shown that Gaussian Wigner functions on the plane may also be simulated efficiently on a classical computer \cite{mari_positive_2012}.  Hudson's theorem established the equivalence between Gaussianity and nonnegativity, and so this result is also a statement about Wigner non-negative states \cite{hudson_when_1974}.  Very recently, a resource theory of non-Gaussianity was proposed to explore how to harness quantum advantages \cite{albarelli_resource_2018}.  Experimental procedures in phase space tomography are also well-established for continuous variable systems \cite{vogel_determination_1989, smithey_measurement_1993, leibfried_experimental_1996,poyatos_motion_1996} as well as spin systems \cite{mcconnell_entanglement_2015}. These studies show the utility of Wigner negativity for analyzing the quantum nature of states and for quantifying resources required for a quantum computing speed-up \cite{veitch_negative_2012}.

Previous work has mainly focused on Heisenberg-Weyl symmetry where the associated phase space is the plane or toroidal lattice, corresponding to the continuous and discrete variants respectively \cite{vourdas_quantum_2004, siyouri_negativity_2016, taghiabadi_revealing_2016, ziane_direct_2018}.  Spin systems correspond to a different symmetry group, SU(2). Hence they are represented on a different phase space related to the Stratonovich-Weyl correspondence: the sphere.  QPDs are well-defined on the sphere \cite{varilly_moyal_1989, agarwal_relation_1981, dowling_wigner_1994, tilma_wigner_2016, koczor_continuous_2020} but not much has been systematically studied about their negativity properties in a manner similar to the planar case \cite{klimov_generalized_2017}.

Here we study Wigner negativity for several common spin states.  We first consider spin coherent states, which are the spin analogs of coherent states on the plane.  While spin coherent states are considered the most classical spin states, their Wigner functions still take negative values and exhibit oscillations around zero, unlike  planar coherent states.  The differences perhaps arise from the compact spherical phase space for finite spin values; the Wigner negativity of these states decreases with increasing spin as they approach the planar coherent states in the infinite-spin limit. 

We next consider GHZ  and N00N states; both are ubiquitous resources for quantum information processing and quantum metrological applications \cite{hillery1999quantum, dowling2008quantum, pezze_quantum_2018}.  While these states, particularly the GHZ state, are usually described in terms of qubits (spin-1/2 systems), they can be equivalently represented using the collective spin of the qubits as superpositions of spin coherent states.  We derive an exact expression for the Wigner function of such superposition states as well as a bound on their Wigner negativity. Our analysis shows that GHZ states and N00N states of equal spin have the same Wigner negativity, and the negativity increases with increasing spin (more qubits).

However for a given collective spin quantum number $j$, they are not the states with the maximum Wigner negativity. We show this by calculating the Wigner negativity of Dicke states, which are the eigenstates $\ket{j,m}$ of the $J_z$ spin operator. The spin coherent states are a special case $\ket{j,\pm j}$ of the Dicke states. We show that most Dicke states $\ket{j,m}$ have greater Wigner negativity than GHZ states with the same value of $j$. We also disprove a conjecture that the Wigner function of the state $\ket{j,m}$ has $2j$ roots by providing a counterexample. Furthermore, we identify which Dicke states have maximal Wigner negativity as the spin $j$ is increased.  While the extremal Dicke states $\ket{j,j}$ have the least Wigner negativity, surprisingly the states $\ket{j,0}$ are not always the most negative.  We explore this behaviour by considering how the Dicke states approach the harmonic oscillator number states in the large spin limit.  Our results show that the negativity depends on the nonclassical nature of the states as well as the underlying structure of the classical phase space.

In section \ref{sec:level2}, we review the Stratonovich-Weyl framework for constructing Wigner functions in the cases of continuous Heisenberg-Weyl and SU(2) symmetry. In section \ref{sec:level3}, we analyse three classes of common spin states: spin coherent, spin cat, and Dicke states.  We compute either numerically or analytically their Wigner functions and Wigner negativities, focusing on changes with respect to increasing spin, and how this compares to the planar scenario.  We end with a discussion, and comment on the differences between Wigner negativity and entanglement, as seen through entanglement entropy and the geometric measure.

\section{\label{sec:level2}Stratonovich-Weyl correspondence}

In this section we review the Stratonovich-Weyl correspondence, and its associated Wigner functions.  More detail may be found in \cite{klimov_generalized_2017, arecchi_atomic_1972, brif_phase-space_1999, varilly_moyal_1989,heiss_discrete_2000}.

\subsection{\label{sec:level2a}Continuous Heisenberg-Weyl}
Specific to a system's dynamical symmetry group, the Stratonovich-Weyl correspondence is realized by an operator-valued distribution over the associated phase space \cite{stratonovich_distributions_1956, brif_phase-space_1999}.  These phase-point operators, collectively called the \textit{kernel} $\Delta(\Omega)$, define the Wigner function of a quantum state $\rho$ by the expectation values
\begin{equation}
    W_\rho (\Omega) = \Tr[\rho \Delta(\Omega)]
\end{equation}
where $\Omega$ is a point in phase space.  The operators $\Delta(\Omega)$ obey several reasonable constraints. The most important one gives the group equivariance condition, ensuring the equivalence between transformations in Hilbert space and phase space:
\begin{equation}\label{eq:general-covariance}
    W_{\pi_g A\pi^\dagger_g}(\Omega) = (g \cdot W_A)( \Omega).
\end{equation}
Here $\pi$ is an irreducible unitary representation of the symmetry group
$\mathcal{G}$, and the action on phase space functions is induced by the action on phase space itself $(g \cdot W_A)( \Omega) := W_A(g^{-1}\cdot \Omega)$ \cite{klimov_generalized_2017}, where $g$ is an element of $\mathcal{G}$.

In the case of continuous Heisenberg-Weyl symmetry in one spatial dimension, i.e.\ $\bs{x,p}=i\id$, the phase space is the complex plane \cite{brif_phase-space_1999}.  This noncompact phase space describes many physical scenarios, including a bosonic field mode and a nonrelativistic spinless particle. 

Defining the
 displacement and parity operators  through their action on the annihilation operator:
\begin{equation}
\begin{aligned}
    D(\alpha)aD(\alpha)^\dagger &= a - \alpha \\
    \Pi a \Pi &= -a
\end{aligned}
\end{equation}
the $H_1(\mathbb{R})$-kernel takes the form of a family of displaced parity operators
\begin{equation}\label{eq:HW-kernel}
    \Delta(\alpha) = 2D(\alpha) \Pi D(\alpha)^\dagger
\end{equation}
where $\alpha = (q+ip)/\sqrt{2}$ is a point in phase space \cite{royer_wigner_1977}.  
The parity operator can also be seen as a phase-shifting operator $\Pi = e^{i\pi N}$ by a rotation $\pi$ about the origin, where $N$ is the number operator \cite{amiet_contracting_2000}. The above definition reduces to the more familiar equation for an $H_1(\mathbb{R})$-covariant Wigner function \cite{royer_wigner_1977}
\begin{equation}\label{eq:common-HW-Wigner}
    W_{\psi}(r,p) = 2\int_{-\infty}^\infty \de s \, e^{-2ips/\hbar} \psi(r - s) \psi^*(r + s)
\end{equation}
for conjugate position and momentum variables $(r,p)$.  

In general, the Wigner negativity $\delta(\rho)$ of a quantum state $\rho$ is
\begin{equation}\label{wigneg}
    \delta(\rho) = \frac{1}{2}\int_{\Gamma} |W_{\rho}(\Omega)| \de \mu(\Omega) - \frac{1}{2} \geq 0
\end{equation}
where $\de\mu(\Omega)$ is the invariant measure on phase space. Throughout this paper the phrase ``Wigner negativity" shall refer to an integrated volume, and is given by a nonnegative real number.

\subsection{\label{sec:level2b}Spin systems}
Quantum spin systems are described by a quantized angular momentum vector $J = (J_x, J_y, J_z)$ of fixed length $|J|=j$, where $j>0$ is an integer or a half-integer.  States living in a $2j+1$ dimensional Hilbert space are acted upon with irreducible unitary representations of SU$(2)$, with size indexed by $j$.  The SU$(2)$ generators $J_i$ satisfy $\lbrack J_i,J_j \rbrack = i\epsilon_{ijk} J_k$ and yield rotation unitaries through their exponentiation.  The eigenstates of any convex combination $n_xJ_x + n_yJ_y + n_zJ_z = \bm{n} \cdot \bm{J}$ form a basis with respect to the axis $\bm{n}$, denoted $\bc{\ket{j,m;\bm{n}}}$. These basis vectors are called Dicke states, and the projection eigenvalue $m$ runs from $-j$ to $j$ in integer steps \footnote{Dicke states are also called spin number states as they are the spin analogues to the number states.}.  In spherical coordinates the vector $\bm{n}=(\sin\theta\cos\phi,\sin\theta\sin\phi,\cos\theta)$ points to $(\theta,\phi)$ on the sphere; this is the phase space generated by SU(2) symmetry \cite{klimov_generalized_2017}.  It is common to work entirely in the $\bm{n}_z$ Dicke basis, denoted $\bc{\ket{j,m}}$.  

The SU$(2)$-kernel at the point $(\theta,\phi)$ is diagonal in the Dicke projector basis along $\bm{n}$ \cite{heiss_discrete_2000}:
\begin{equation}\label{eq:su2-kernel-diagonal}
\Delta_j(\theta,\phi) = \sum_{m=-j}^j \Delta_{j,m} \ketbra{j,m;\bm{n}}{j,m;\bm{n}},
\end{equation}
with eigenvalues
\begin{equation}\label{eq:kernel_eigenvalues}
\Delta_{j,m} = \sum_{l=0}^{2j} \varepsilon_l \frac{2l+1}{2j+1} \cg{j}{m}{l}{0}{j}{m}.
\end{equation}
Here $\varepsilon_0 =1$, $\varepsilon_l = \pm 1$, and $\cg{j_1}{m_1}{j_2}{m_2}{J}{M}$ are Clebsch-Gordan coefficients.  See also \cite{varilly_moyal_1989, klimov_generalized_2017, koczor_continuous_2020} for different characterizations of the SU(2)-kernel.

There are $2^{2j}$ choices of $\varepsilon_l$ leading to different eigenvalue distributions, indicative of  the non-uniqueness of valid SU$(2)$-kernels \cite{varilly_moyal_1989}.  In the limit of infinite spin however, only one such choice leads to a natural contraction from the spherical kernel to the planar kernel, namely $\varepsilon_l = 1$ for all $l$  \cite{amiet_contracting_2000}.  Intuitively seen as a tangent plane to a sphere of increasing radius, this choice naturally connects the two types of displacements and parities defined by their symmetry (e.g.\ rotations become translations) \cite{amiet_contracting_2000}.  As discussed in section \ref{sec:dicke}, this contraction also sees the number state as a limit of the Dicke state:
\begin{equation}\label{eq:number-state-relation}
    \lim_{j\rightarrow \infty} \ket{j,j-n} = \ket{n}
\end{equation}
for fixed $n$ \cite{arecchi_atomic_1972}.  

Picking this choice of $\varepsilon_l$, the SU(2)-kernel, similar to the Heisenberg-Weyl kernel \eqref{eq:HW-kernel}, is also realized as a family of displaced parity operators:
\begin{equation}\label{eq:SU(2)-kernel}
    \Delta_j(\theta,\phi) = U_{\bm{k}} \Delta_j(0,0) U_{\bm{k}}^\dagger,
\end{equation}
where $(0,0)$ is the north pole and $U_{\bm{k}} = e^{-i\theta \, \bm{k}\cdot\bm{J}}$ is the unitary taking the north pole to $(\theta, \phi)$ by a rotation about $\bm{k} = (-\sin\phi, \cos\phi, 0)$.  Using this kernel, the SU(2)-covariant Wigner function of a quantum state $\rho$ is
\begin{equation}
    W_{\rho}(\theta,\phi) := \Tr[\rho\Delta_j(\theta,\phi)],
\end{equation}
where $j$ matches the representation size.

\section{\label{sec:level3}Spin states}

In this section, we analyse the Wigner functions and Wigner negativities of spin coherent states, spin cat states, and Dicke states.  We compare our findings with the Heisenberg-Weyl analogues on the plane.

\begin{figure}[h!]
    \centering
    \includegraphics[width=\columnwidth]{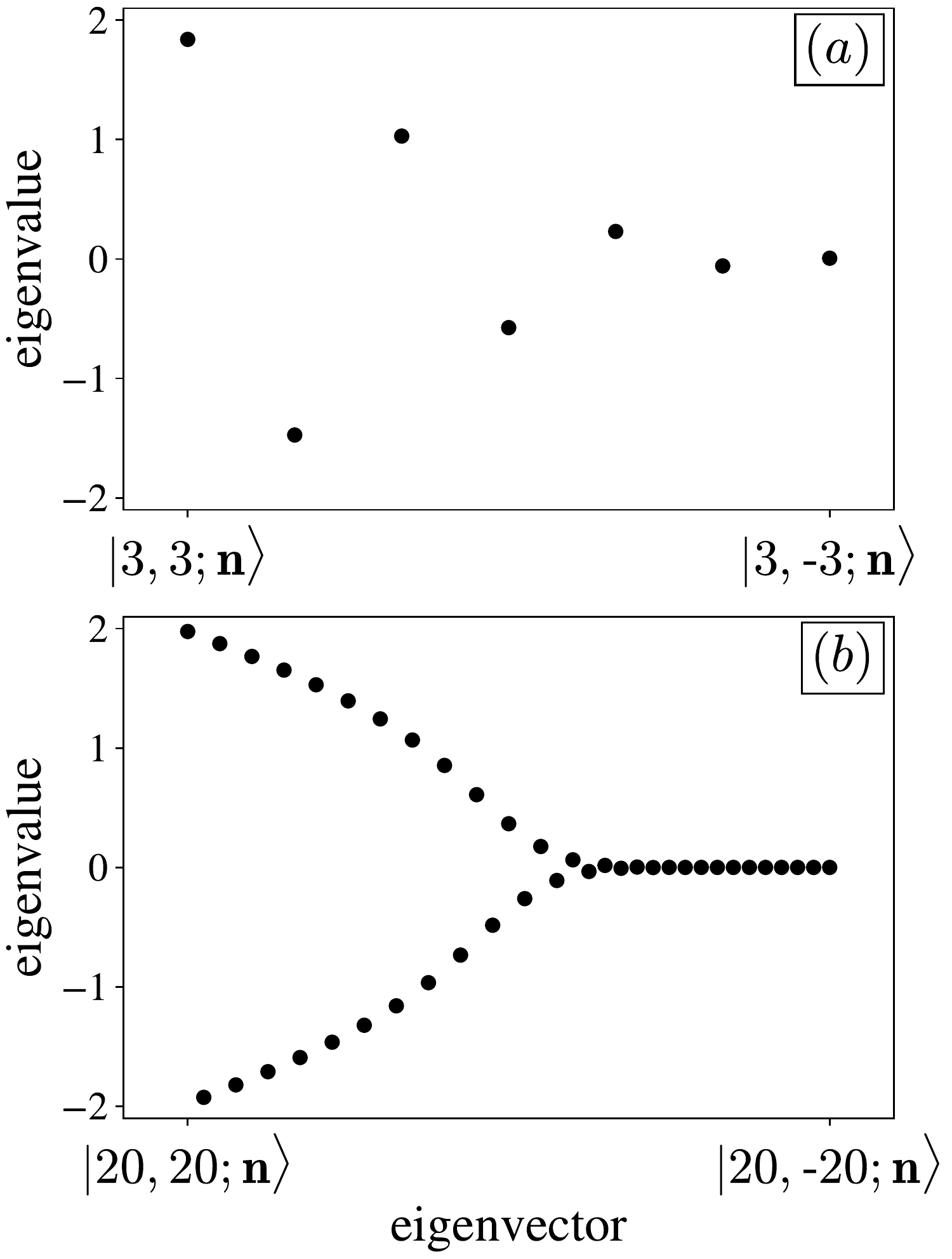}
    \caption{Spectrum of the SU(2) kernel in Eq.\ \eqref{eq:kernel_eigenvalues} at any point $\Omega \in S^2$ for spins (a) $j=3$ and (b) $j=20$.  For low spin the maximum and minimum eigenvalues are not equal in absolute magnitude but approach $\pm 2$ as spin increases, matching the planar Wigner function bounds.}
    \label{fig:kernel-eigenvalues}
\end{figure}

But before doing so, we briefly give a simple and motivating example of how the planar and spherical Wigner functions have different global properties.  In particular, their pointwise upper and lower bounds are distinct.  We have plotted in Fig.\ \ref{fig:kernel-eigenvalues} the eigenvalue distribution of the SU$(2)$-kernel \eqref{eq:kernel_eigenvalues} for low and high spin. For spherical Wigner functions, the maximum eigenvalue of the kernel in Eq.\ \eqref{eq:su2-kernel-diagonal} gives the pointwise upper bound while the minimum eigenvalue gives the pointwise lower bound.  Since the eigenvalues $\Delta_{j,m}$ are independent of the quantization axis $\bm{n}$, we see from Fig.\ \ref{fig:kernel-eigenvalues} that the pointwise upper bound is attained by the eigenstate $m=j$ along $\bm{n}$, corresponding to a spin coherent state centered at $(\theta,\phi)$.  The pointwise lower bound corresponds to the eigenstate $\ketbra{j,j-1;\bm{n}}{j,j-1;\bm{n}}$, which as discussed in section \ref{sec:dicke} is incidentally the $\textsf{W}$ state.  We see that the pointwise upper bound of the spherical Wigner function is larger in absolute value than its pointwise lower bound. In contrast, the pointwise upper and lower bounds of the planar Wigner function have the same absolute value of 2; see Eq.\ \eqref{eq:HW-kernel}. In the case of single qubit states (i.e.\ $j=1/2$), the pointwise bounds are
\begin{equation}
    \begin{split}
        \max\bs{ W_{\text{qubit}}(\theta,\phi) } &= \frac{1}{2}\bp{1 + \sqrt{3}} \approx 1.37 \\
        \min\bs{ W_{\text{qubit}}(\theta,\phi) } &= \frac{1}{2}\bp{1 - \sqrt{3}} \approx -0.37.
    \end{split}
\end{equation}
In the limit of infinite spin, these converge to $\pm 2$, matching those of the planar Wigner function \cite{amiet_contracting_2000}.

\subsection{Spin coherent states \label{SEC:scs}}

Similar to planar coherent states, spin coherent states may be defined as the displacement of some reference state. A commonly chosen reference state is $\ket{j,j}$, whose Wigner function is centred at the north pole of the sphere. The spin coherent state after the displacement (i.e.\ rotation) can be written as:
\begin{equation}
\begin{aligned}
    \ket{\theta, \phi} &= U_{\bm{k}}\ket{j,j} \\ &= e^{-i\theta \, \bm{k}\cdot \bm{J}} \ket{j,j} \\
    &= e^{-\frac{\theta}{2}\bp{J_+ e^{-i\phi} - J_- e^{i\phi}} }\ket{j,j}
\end{aligned}
\end{equation}
where $J_{\pm}$ are the spherical ladder operators \cite{klimov_generalized_2017} and $\bm{k} = (-\sin\phi, \cos\phi,0)$ is the axis of rotation with $0 \leq \theta \leq \pi $, $0 \leq \phi < 2\pi$.

The Wigner negativity of a spin coherent state, and for spin states in general, can be calculated using Eq.\ \eqref{wigneg} where $\de\mu(\Omega) = \frac{2j+1}{4\pi} \sin\theta\de\theta\de\phi$:
\begin{equation}\label{eq:wig_neg_SU2}
    \delta(\rho) = \frac{1}{2} \left( \frac{2j+1}{4\pi}\int_{\theta=0}^{\pi}\int_{\phi=0}^{2\pi} |W_{\rho}(\theta,\phi)| \sin{\theta} \de \theta \de \phi - 1 \right)
\end{equation}
\footnote{The $(2j+1)/4\pi$ factor ensures the identity operator is matched with the unit function on $S^2$.}.  Since this measure is invariant under global rotations, states connected through global rotations will have the same Wigner negativity.  Spin coherent states are intrinsically defined this way and so we focus on the state $\ket{j,j}$ as a representative:
\begin{equation}
\begin{aligned}
    & W_{\ket{j,j}}(\theta,\phi) = \\
    &\,\, \frac{(2j)!}{\sqrt{2j+1}} \sum_{l=0}^{2j} \frac{(2l+1)}{\sqrt{(2j-l)!(2j+1+l)!}} P_l(\cos\theta),
\end{aligned}
\end{equation}
where $P_l(\cdot)$ is the $l$-th Legendre polynomial \cite{varilly_moyal_1989}.

\begin{figure}[h!]
    \centering
    \includegraphics[width=\columnwidth]{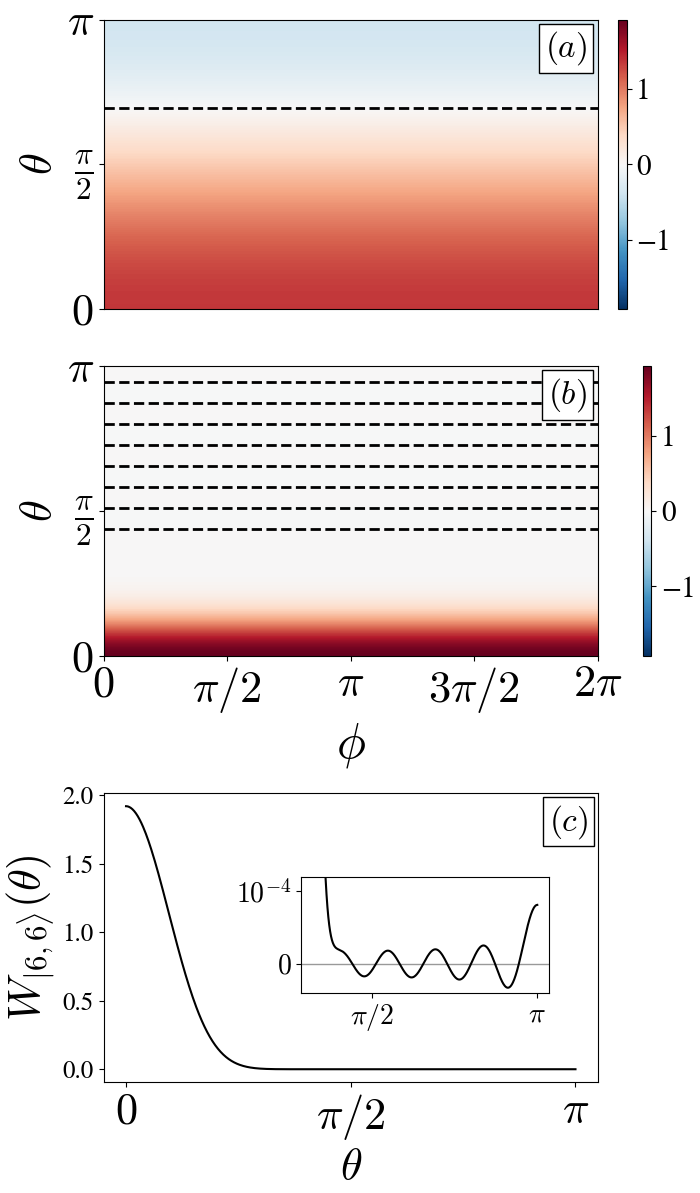}
    \caption{Wigner function of the spin coherent state $\ket{j,j} = \ket{\theta_0=0}$ associated with a spin of (a) $j=1/2$ (single qubit) and (b) $j=6$ (twelve qubits).  Figure (c) shows the polar cross section of the $j=6$ spin coherent state where the inset illustrates the roots of the associated Wigner function.  Dashed lines in (a) and (b) are along roots.}
    \label{fig:scs-0.5-and-5}
\end{figure}

The Wigner function of optical coherent states is Gaussian in the field quadratures and so may only take positive values.  In contrast, the Wigner function of spin coherent states has nonzero Wigner negativity (Fig.\ \ref{fig:scs-0.5-and-5}).  For example, a single spin-1/2 system (a qubit) prepared in the state $\ket{0} = \ket{1/2,1/2}$ has the Wigner function
\begin{equation}\label{eq:qubit-spins-wigner}
        W_{|\frac{1}{2},\frac{1}{2}\rangle} (\theta,\phi) = \frac{1}{2} + \frac{\sqrt{3}}{2}\cos{\theta}.
\end{equation}
Analyzing Eq.\ \eqref{eq:qubit-spins-wigner}, we find the negativity of this state to be $\frac{1}{2} - \frac{1}{\sqrt{3}} \approx 0.077$, matching the results from \cite{arkhipov_negativity_2018}.  It follows that all pure single qubit states share this value.  As discussed in \cite{varilly_moyal_1989}, qubit states must be sufficiently mixed in order for their Wigner function to be nonnegative.  In particular, their Bloch vector must have magnitude less than $1/\sqrt{3}\approx 0.58$, defining an inner Bloch ball of Wigner-positive states.

As the spin $j$ is increased, we have numerically confirmed that the Wigner negativity of a spin coherent state rapidly approaches zero, although it does not vanish for the finite $j$ considered ($j<80$).  All of the negativity  contributions come from small oscillations in the Wigner function, generally present in the hemisphere opposite the centroid $(\theta,\phi)$; see Fig.\ \eqref{fig:scs-0.5-and-5}.  In planar phase space, such oscillations are usually associated with a superposition of distinct macroscopic states (e.g.\ a planar cat state), which is highly nonclassical. However, the spin coherent state is typically considered the most classical-like spin state because of its analogy to Gaussian coherent states on the plane.  Thus Wigner negativity helps identify the important differences between planar and spin coherent states that exist despite their similarities.  The nonvanishing negativity for finite spin seems to result from the compact spherical phase space compared to the infinite planar phase space of Gaussian coherent states. 

We also find a surprising result that refutes a conjecture in \cite{varilly_moyal_1989} proposing that all spin $j$ Dicke state Wigner functions (including spin coherent states) have $2j$ distinct roots.  We numerically give a counterexample: the spin coherent state of 12 qubits has only 8 distinct roots as illustrated in Fig.\ \ref{fig:scs-0.5-and-5}.

In the limit of infinite spin, the tangent plane to a sphere SU(2)-kernel contraction implies that the spin coherent state approaches the planar coherent state.  Indeed, this is the special case of Eq.\ \eqref{eq:number-state-relation} for $n=0$, demonstrating that, up to displacements, the state $\ket{j,j}$ approaches the harmonic oscillator vacuum, and so the corresponding Wigner negativities approach zero as seen in the $n=0$ case in Fig.\ \ref{fig:comparison}.

\subsection{Spin cat states}\label{sec:ghz}

The Greenberger-Horne-Zeilinger (GHZ) state was first introduced to generalize Bell's theorem on quantum nonlocality to a multipartite setting \cite{greenberger_going_1989}. A closely related set of states called N00N states were introduced in the same year for their use in understanding decoherence of cat-like states \cite{sanders_quantum_1989}.  Both of these have since been intensively studied within quantum information, quantum optics, and quantum metrology \cite{hillery1999quantum, dowling2008quantum, pezze_quantum_2018}.

The GHZ and N00N states, together with their higher multipartite generalizations, can be naturally viewed as elements of a spin system, with both taking the general form $\alpha\ket{j,j;\bm{n}} + \beta\ket{j,-j;\bm{n}}$.  The GHZ state is typically defined within the symmetric qubit realization of spin systems:
\begin{equation}\label{eq:subspace-realization}
    J_i = \frac{1}{2} \sum_{l=1}^{2j} \sigma^{(l)}_i
\end{equation}
where $\sigma^{(l)}_i$ denotes the $i$-th Pauli operator of the $l$-th qubit.  In this qubit representation, the computational states $\ket{0}^{\otimes N}$ and $\ket{1}^{\otimes N}$ are identified with the north and south pole spin coherent states $\ket{j, \pm j}$ respectively.  With this, the $N$-qubit GHZ state is given by
\begin{align}
    \ket{\text{GHZ}^{(N=2j)}} &= \frac{1}{\sqrt{2}}(\ket{0}^{\otimes N} + \ket{1}^{\otimes N}) \nonumber \\
    &= \frac{1}{\sqrt{2}}( \ket{j,j} + \ket{j,-j} ).
    \label{eq:GHZ}
\end{align}

But the $\mathfrak{su}(2)$ algebra can also be realized optically using a fixed number of $2j$ photons distributed over two distinct modes, sometimes referred to as the Schwinger realization:
\begin{equation}\label{eq:Schwinger-realization}
    J_+ = a^\dagger b \qquad
    J_- = b^\dagger a \qquad
    J_z = \frac{1}{2}\bp{ a^\dagger a - b^\dagger b }
\end{equation}
where $a$ ($a^\dagger$) and $b$ ($b^\dagger$) are the annihilation (creation) operators of the respective $a$ and $b$ modes  \cite{sanders_connection_2014}.  With the two modes playing the role of collective spin-up and spin-down, optical Dicke states are
\begin{equation}\label{eq:schwinger-realization}
    \ket{j, m} = \ket{j + m}_a \ket{j - m}_b.
\end{equation}
The N00N state is defined similarly to the GHZ state but with an additional spin-dependent relative phase:
\begin{align}
    \ket{\text{N00N}^{(N=2j)}} &= \frac{1}{\sqrt{2}} ( \ket{N}\ket{0} + e^{iN\theta}\ket{0}\ket{N} ) \nonumber \\
    &= \frac{1}{\sqrt{2}} ( \ket{j,j} + e^{i2j\theta}\ket{j,-j} ).
\end{align}
From the phase space perspective, the Schwinger realization stems from placing a restriction on the tensor product of two previously existing Heisenberg-Weyl systems, while the symmetric qubit realization comes directly from the induced irreducible representations over the symmetric tensor power of a single qubit Hilbert space.

\begin{figure}[h!]
    \centering
    \includegraphics[width=\columnwidth]{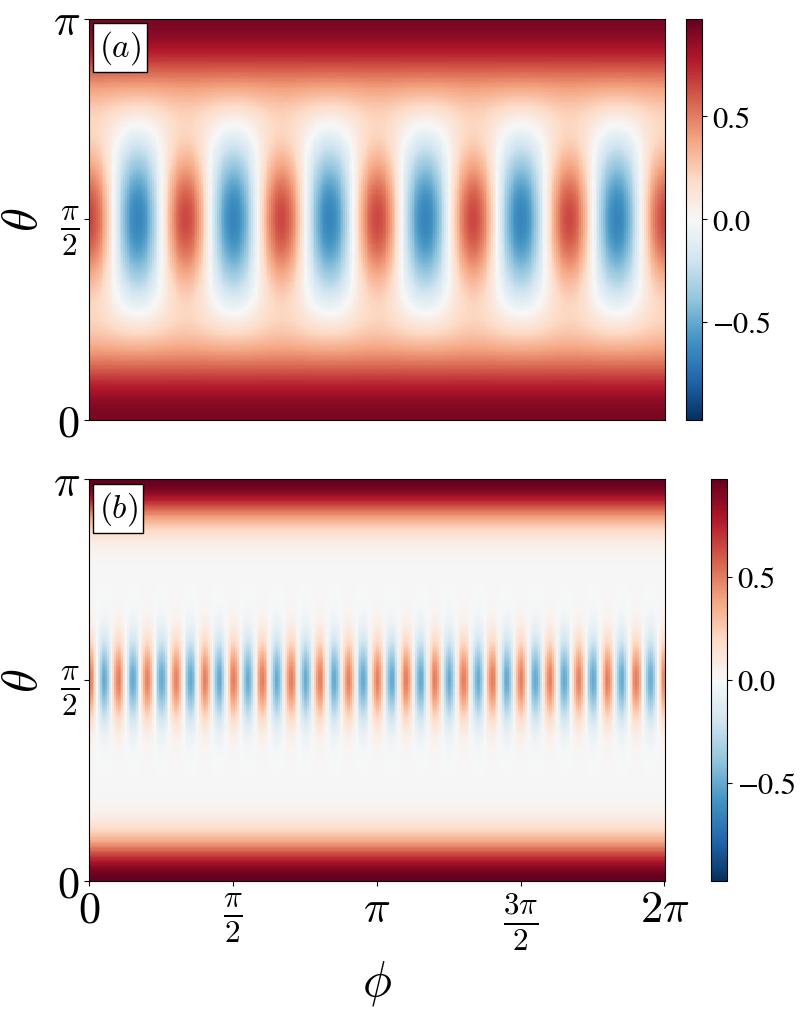}
    \caption{Wigner function of the GHZ state corresponding to (a) 6 qubits ($j=3$) and (b) 20 qubits ($j=10$).}
    \label{fig:ghz-j10}
\end{figure}

Here, we consider the general spin cat state
\begin{equation}\label{eq:g-cat-state}
    \ket{j,\vartheta,\varphi} = \cos\bp{\frac{\vartheta}{2}} \ket{j,j} + e^{i\varphi}\sin\bp{\frac{\vartheta}{2}}\ket{j,-j}
\end{equation}
where $0 \leq \vartheta \leq \pi$ and $0 \leq \varphi \leq 2\pi$.  
\begin{widetext}
We derive the exact Wigner function of this state to be
\begin{equation}\label{eq:generalized-ghz-noon-wigner-fnx}
    W_{\ket{j,\vartheta,\varphi}}(\theta, \phi) = \cos^2\frac{\vartheta}{2} \, \, W_{\ket{j,j}}(\theta,\phi) + \sin^2 \frac{\vartheta}{2} \, \, W_{\ket{j,-j}}(\theta,\phi) + \sin\vartheta \,\, N_j \sin^{2j}(\theta) \cos(2j\phi - \varphi)
\end{equation}
\end{widetext}
where
\begin{equation}
    N_j = \frac{1}{2^{2j} (2j)!} \sqrt{\frac{(4j+1)!}{2j+1}}
\end{equation}
as we show in appendix \ref{sec:wigner-analytics}.  The first two terms of Eq.\ \eqref{eq:generalized-ghz-noon-wigner-fnx} correspond to the weighted Wigner function of two antipodal spin coherent states.  The interference pattern from the third term is expressed throughout phase space as a band of fringes along the equator, with the number of negative islands equal to the $2j$ number of qubits.  As spin increases, the interference pattern becomes more concentrated along the equator, while the spatial extent of the positive polar regions shrinks. This is a consequence of the polar regions locally approaching that of a planar coherent state.  This, in addition to the $\sin^{2j}(\theta)$ factor in the interference term, highly suppresses the Wigner function in the regions between the equator and the two poles; see Fig.\ \eqref{fig:ghz-j10} for $j=3$ and $j=10$ in the GHZ case of $(\vartheta, \varphi) = (\pi/2,0)$.
 
\begin{figure}
    \centering
    \includegraphics[width=\columnwidth]{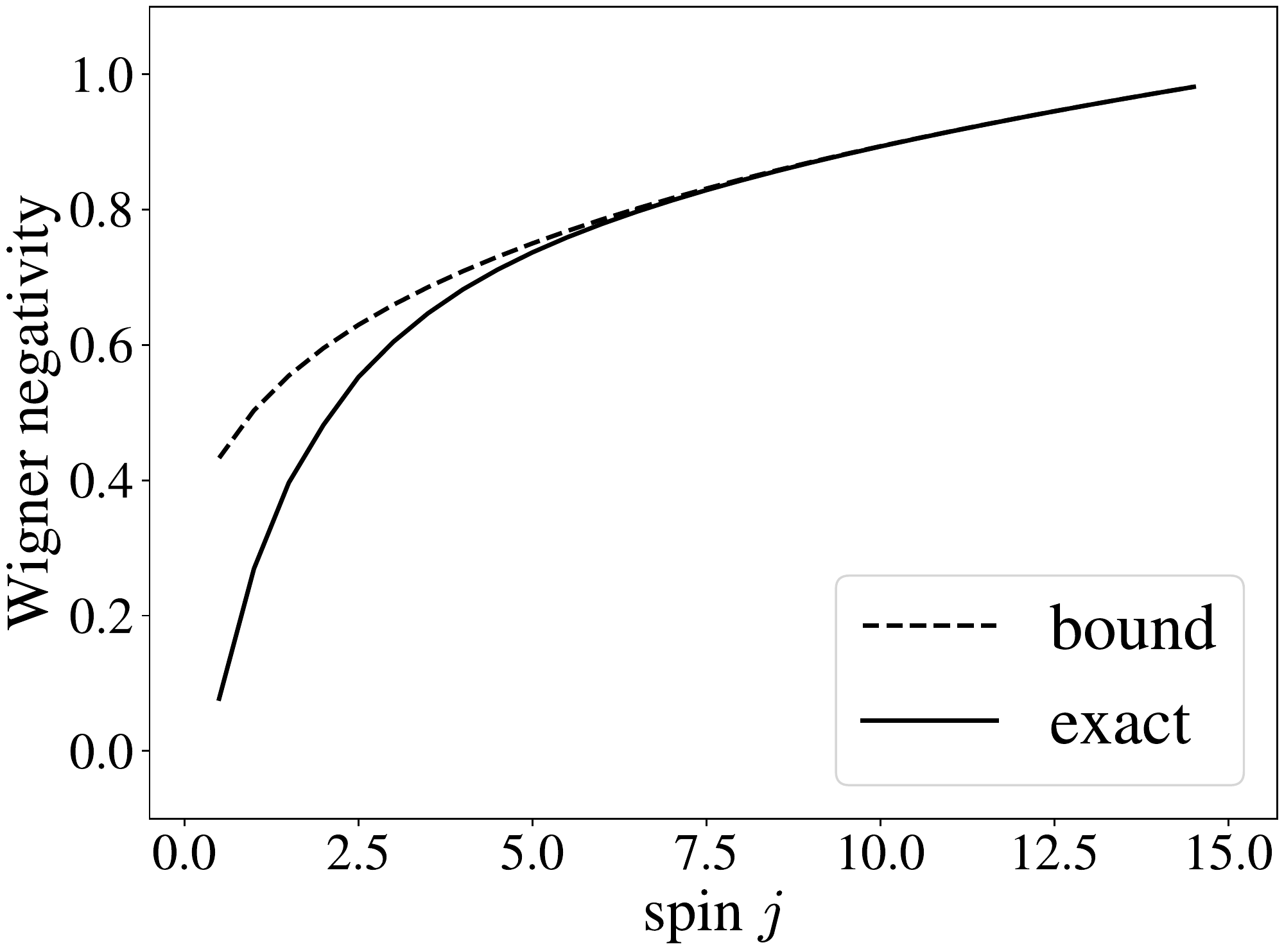}
    \caption{Comparison between our spin cat state Wigner negativity bound \eqref{eq:ghz-general-neg} and the exact Wigner negativity.}
    \label{fig:ghz_neg_compare}
\end{figure}

The interference term is the primary contribution to the Wigner negativity of the state due to the rapidly vanishing contributions from the spin coherent components.  By focusing exclusively on the interference fringes, we derive the upper bound on the Wigner negativity of these states as
\begin{align}\label{eq:ghz-general-neg}
    \sin \vartheta N_j \frac{\sigma(j)}{\pi} \frac{(2j)!!}{(2j-1)!!}
\end{align}
where $\sigma(j) = 1$ for integer spin and $\sigma(j) = \pi/2$ for half-integer spin; see appendix \ref{sec:wigner-analytics} for details.  The explicit Wigner negativity $\delta(\cdot)$ for these two cases are:
\begin{equation}\label{eq:ghz-general-neg2}
\begin{aligned}
    \delta^{\text{(int)}} &\lesssim \frac{1}{\pi} \sin\vartheta \sqrt{\frac{(4j+1)!}{2j+1}} \bp{ \frac{j!}{(2j)!} }^2 \\
    \delta^{\text{(half-int)}} &\lesssim \sin\vartheta \sqrt{\frac{(4j+1)!}{2j+1}} \frac{1}{2^{4j}} \frac{1}{(j-\frac{1}{2})!^2}.
\end{aligned}
\end{equation}
In Fig. \ref{fig:ghz_neg_compare}, we have plotted the exact Wigner negativity together with the bound as a function of $j$. This bound provides a good estimate of the Wigner negativity for $j \gtrsim 5$.

In the N00N states, a relative phase $e^{i\varphi}$ between the two spin coherent states amounts to a global rotation about the quantization axis connecting them; the fringes along the equator are shifted by an amount $\varphi$, implying that the GHZ and N00N states have the same Wigner negativity.  If there is an asymmetric weighting in the superposition \eqref{eq:g-cat-state} parameterized by $\vartheta$, the interference fringes, and consequently our lower bound, is suppressed by a factor of $\sin\vartheta$ for all spin.
\begin{figure}[h!]
    \centering
    \includegraphics[width=\columnwidth]{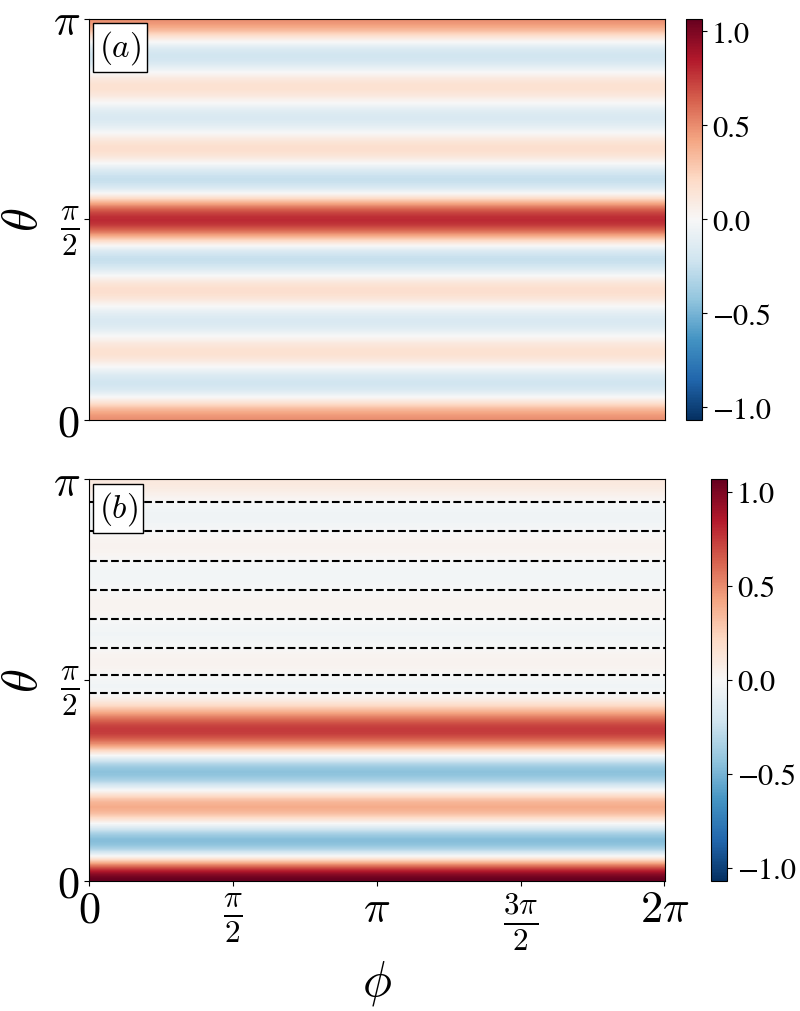}
    \caption{Wigner functions of the Dicke states (a) $\ket{6,0}$, and (b) $\ket{6,2}$.  Dashed lines mark difficult to see roots of the Wigner function.  The non-equatorial state in (b) displays reduced amplitude in the southern hemisphere region.}
    \label{fig:dicke-wigner-examples}
\end{figure}

\subsection{Dicke states}\label{sec:dicke}
Having analyzed spin coherent states $\ket{j,\pm j}$ and superpositions of spin coherent states, we now turn our attention to the more general set of Dicke states $\ket{j,m}$.  Like the GHZ and N00N states, these also have a multiqubit representation.
The $\ket{j,m}$ states are generalized $\textsf{W}$-like states, first introduced in the SLOCC classification of multipartite entanglement  \cite{dur_three_2000}.  The multipartite $\textsf{W}$-like state of weight $k \in \bc{0, \cdots, N}$ is the symmetrized superposition of $N = 2j$ qubits where $k$ out of $N$ of them are in the excited state $\ket{1}$:
\begin{equation}\label{eq:W-Dicke-def}
    \ket{D^{(k)}_N} = \binom{N}{k}^{-\frac{1}{2}} \sum_{\tau_i \in S_n} | \tau_i(  \underbrace{1\cdots 1}_{k} \,  \underbrace{0\cdots 0}_{N-k}) \rangle  \leftrightarrow \ket{j,j-k}.
\end{equation}
where $S_n$ is the symmetric group of order $n$.  The case $k=1$ is the standard $\textsf{W}$ state, while the two extremal cases $\ket{j,\pm j}$ correspond to antipodal spin coherent states.  A collective spin-flip of a Dicke state, $\sigma_x^{\otimes N}\ket{D^{(k)}_N}$, is its conjugate state
\begin{equation}\label{eq:conjugate-Dicke-state}
    |\overline{D_N^{(k)} }\rangle := | D_N^{(2j-k)}\rangle \leftrightarrow \ket{j, -j + k}.
\end{equation}
As the $m$ eigenvalue in $|j,m\rangle$ approaches either $\pm j$, the equivalent $\textsf{W}$-like state has an increasingly asymmetric ratio of ground to excited qubits -- i.e.\ mostly 0s or mostly 1s, corresponding here to the northern and southern hemispheres of phase space.

The spherical Wigner function of the Dicke state $\ket{j,m}$ is given by
\begin{align}\label{eq:dicke-state-wigner-function}
    W_{\ket{j,m}}(\theta, \phi) &= \langle j,m | \Delta_j(\theta, \phi) | j,m \rangle \nonumber \\
    &= \sum_{l=0}^{2j} \frac{2l + 1}{2j + 1} \cg{j}{m}{l}{0}{j}{m}
    P_l\bp{\cos\theta}
\end{align}
\cite{varilly_moyal_1989, klimov_generalized_2017}.  We plot this function for two different Dicke states in Fig.\ \ref{fig:dicke-wigner-examples}.  They are characterized by a principle band of positive values around a circle of constant latitude (corresponding to the $m$ projection eigenvalue), with additional alternating bands along the sphere.  If the principle band is distinctly in one hemisphere, the fringes in the opposing hemisphere are reduced in amplitude; see Figs.\ \ref{fig:scs-0.5-and-5} and \ref{fig:dicke-wigner-examples}.  As mentioned earlier, the number of roots for all $m$ values is in general not equal to $2j$.

\begin{figure}[h!]
    \centering
    \includegraphics[width=0.97\columnwidth]{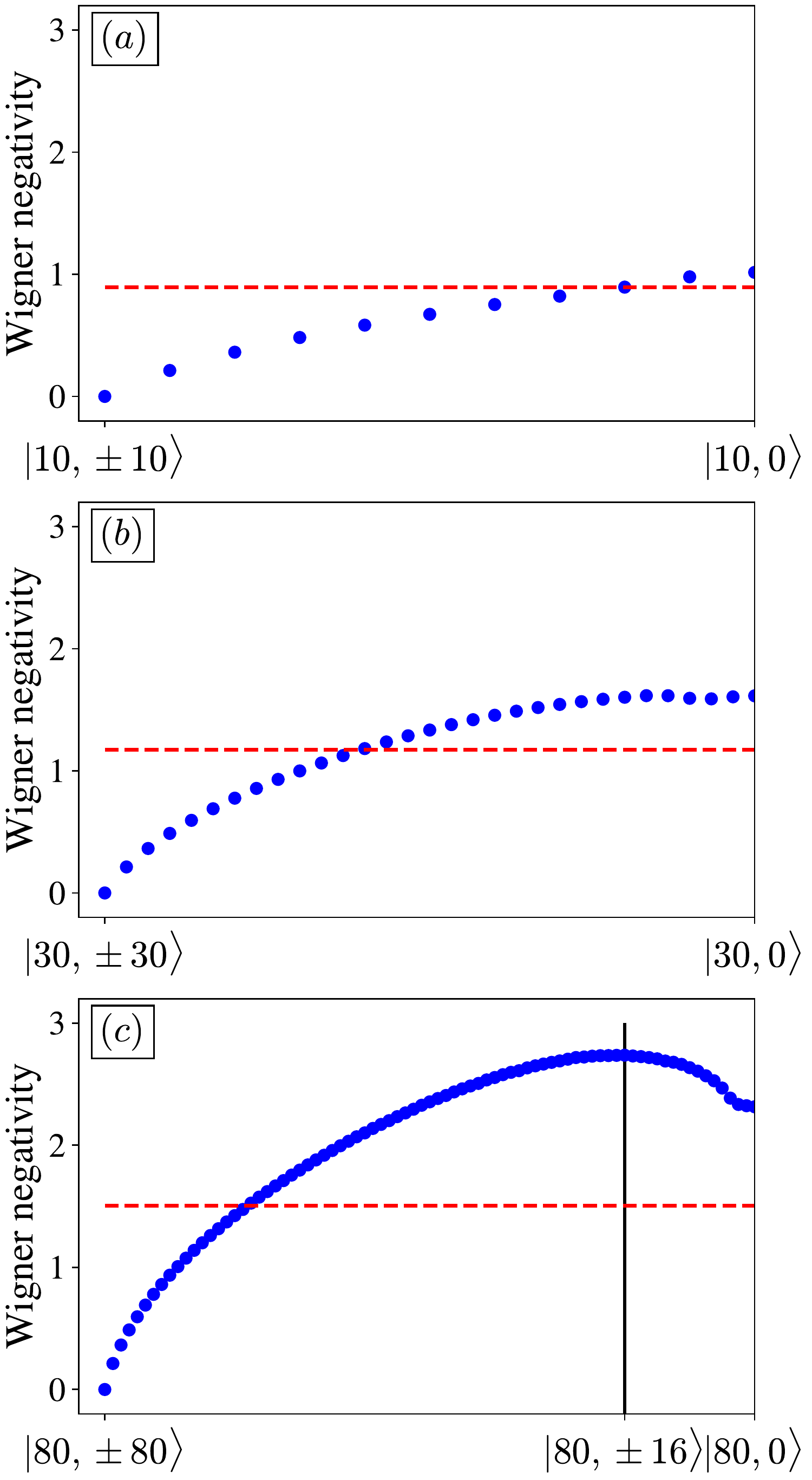}
    \caption{The blue dots are Wigner negativities of the Dicke basis $\bc{\ket{j,\pm m}}$ corresponding to (a) $j=10$, (b) $j=30$, and (c) $j=80$.  The solid vertical line in (c) denotes the maximally Wigner-negative Dicke state.  The red dashed line corresponds to the GHZ states of equal dimension.}
    \label{fig:dicke-and-fock-negativities}
\end{figure}

Next we analyze the Wigner negativity of Dicke states and highlight that it reveals a surprisingly rich structure, dependent on the radius of phase space through the spin $j$.  We begin by focusing on how the Wigner negativity of the entire Dicke basis changes with increasing spin.  See Fig. \ref{fig:dicke-and-fock-negativities} for a few numerical examples of Dicke basis negativities.  We briefly note that conjugate Dicke states within a given Dicke basis have the same negativity; this is because they are related by a global $\pi$ rotation  about any axis in the equatorial plane.  The least negative states of a Dicke basis are always the $\ket{j,\pm j}$ spin coherent states as expected.  We also observe that for fixed $j$, the negativity generally increases as $m$ moves away from the poles, $\pm j$.  This pattern of increasing Wigner negativity with decreasing $|m|$ value continues for low spins $j \lesssim 30$, and culminates with the maximally negative Dicke state lying on the equator: $\ket{j,0}$ for integer spin and $\ket{j,\pm 1/2}$ for half-integer spin.  Surprisingly, this pattern changes for Dicke bases with $j \gtrsim 30$, where the maximally negative state bifurcates away from the equator into a spin-dependent conjugate pair $\ket{j,\pm m'_j}$ with $m'_j > 1/2$; see Fig.\ \ref{fig:dicke-and-fock-negativities}.  For example, the maximally Wigner-negative Dicke state for $j=80$ happens to be $|80,\pm 16 \rangle$ rather than $|80,0 \rangle$; see Fig. \ref{fig:dicke-and-fock-negativities}.  The projection eigenvalue $m'_j$ does not settle to a fixed value for the spins considered (up to $j=80$, or 160 qubits).

\begin{figure}[h!]
    \centering
    \includegraphics[width=\columnwidth]{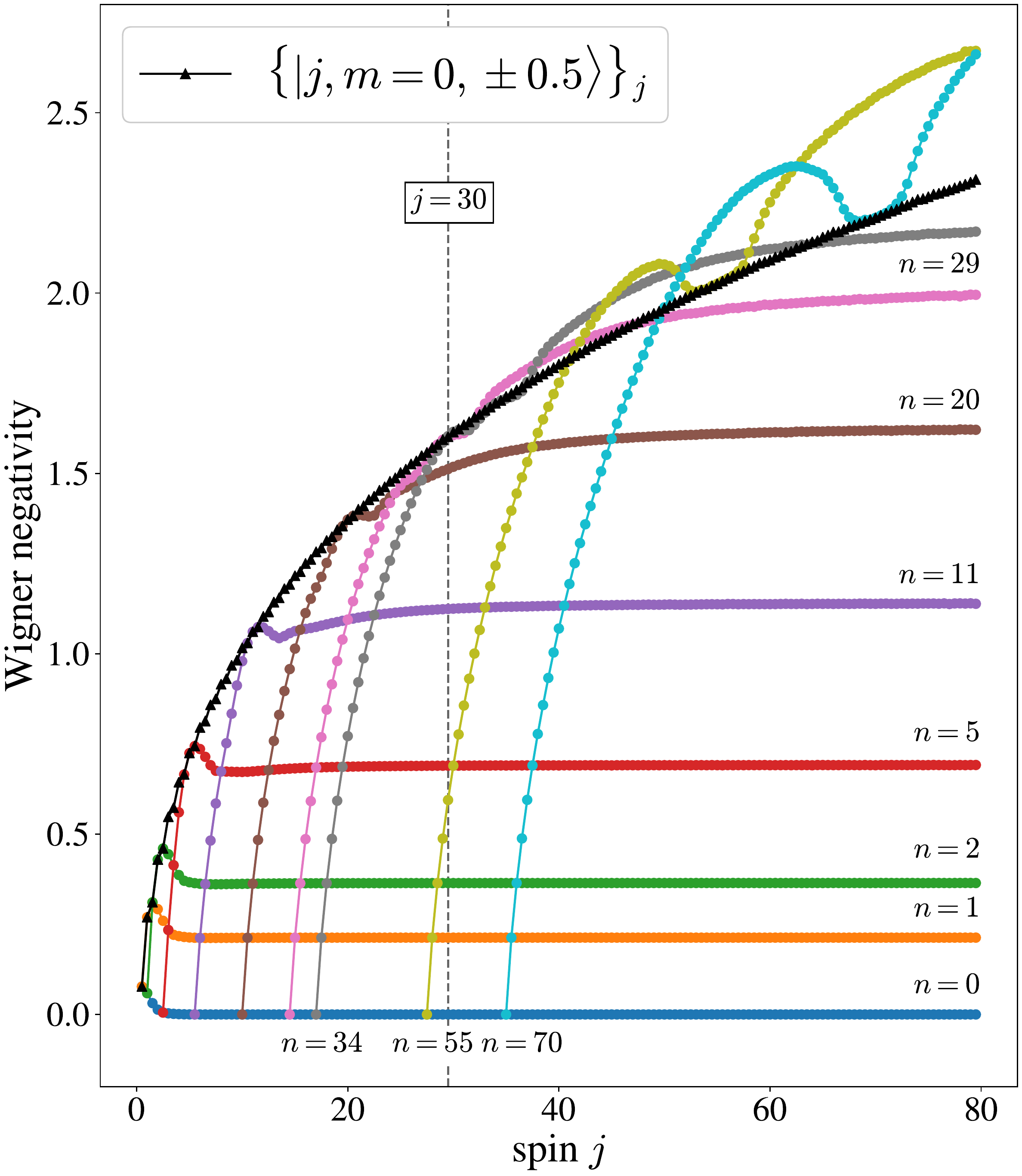}
    \caption{Wigner negativity of sequences of states $\ket{j,j-n}$ as $j$ increases for various fixed values of $n$.  The asymptotic flatlines of a given sequence match the number state negativity $\ket{n}$ as expected.  The black curve with no label corresponds to the equatorial states $\ket{j,0}$ and $\ket{j,\pm 0.5}$.  Around $j \gtrsim 30$ (vertical dashed line) there is emergent behaviour, in which sequences cross over the black equatorial curve. 
    }
    \label{fig:comparison}
\end{figure}

Another way to explore the negativity of Dicke states is to consider their high spin limit towards harmonic oscillator number states $|n\rangle$ \cite{arecchi_atomic_1972}.  To that end, we numerically confirmed Eq.\ \eqref{eq:number-state-relation}, $\ket{j,j-n} \rightarrow \ket{n}$ as $j\rightarrow\infty$, by computing the Wigner negativity of state sequences $\{\ket{j,j-n}\}_j$ for a handful of fixed $n$.  See Fig.\ \ref{fig:comparison} for a collection of results.  To help read this figure, we first point out three aspects common to each sequence corresponding to a fixed $n$ value. The first aspect is that they all begin at some point along the $\{\ket{j,j}\}_j$ curve (i.e.\ when $n=0$).  This is because the starting state for any given $n$, with $n=2j$, is always a spin coherent state on the south pole $\ket{j,-j}$, which has the same negativity as $\ket{j,j}$.  The second is that when $j=n$ or $j=n\pm 0.5$, each sequence is respectively at $\ket{j,0}$ or $\ket{n\pm 0.5, \pm 0.5}$.  This means that each curve will meet the included equatorial curve $\{\ket{j,m=0, \pm 0.5}\}_j$, shown in black, three times (in a row).  The third is that when $j$ is large compared to $n$, each curve asymptotically approaches the negativity of its limiting number state $\ket{n}$; see the visible flatline values.

Despite these common properties, there is clearly nontrivial behaviour occurring as $n$ changes.  For example, low $n$ sequences ($n \lesssim 8$) contain states that are more Wigner-negative than their number state limit.  For $n \gtrsim 8$ this is no longer true.  There is also an emergent behaviour as $n \gtrsim 30$.  In particular, sequences with $n \gtrsim 30$ cross over the equatorial curve.  This is in fact the relationship between Figs.\ \ref{fig:dicke-and-fock-negativities} and \ref{fig:comparison}.  Indeed, consider a vertical cross-section in Fig.\ \ref{fig:comparison}.  This corresponds to a fixed $j$ Dicke basis.  For vertical cross-sections corresponding to $j \lesssim 30$, all spin-$j$ Dicke states lie below the spin-$j$ equatorial state.  On the other hand, for vertical cross-sections corresponding to $j\gtrsim 30$, there are states above the equatorial state, with this effect increasingly exaggerated as $j$ increases.
Thus the aforementioned bifurcation of the most Wigner-negative Dicke state around $j \approx 30$ can also be inferred from Fig.\ \ref{fig:comparison}.

Overall, these two related results of the maximally negative Dicke state being spin-dependent, together with the number state limit display an intriguing structure to the nonclassicality of Dicke states. That this happens for any finite spin is surprising.  Furthermore, the geometric significance of spin suggests a nontrivial interplay between the size (or other properties such as curvature) of phase space and state nonclassicality.

\section{Summary and discussion}

We have investigated the nonclassicality of common spin states as quantified by their Wigner negativity.  We compared our results to the planar scenario and explored the dependence on spin.  Our analysis was based on the Stratonovich-Weyl phase space framework, using the SU(2)-kernel to reproduce Wigner functions of spin states. We analysed spin coherent states, and derived an exact expression for the Wigner function for arbitrary spin-$j$ cat states.  From this, we obtained a bound on the Wigner negativity of spin cat states.  Using the connection between spin states and symmetric ensembles of qubits, we  quantified the Wigner negativity of important qubit states such as the GHZ states and N00N states.  We showed that GHZ and N00N states with the same number of qubits have identical Wigner negativities, but counterintuitively, they may have lower negativities than W-like states of equal dimension.  We also showed that the Wigner function of spin-$j$ Dicke states do not always have $2j$ roots as was previously conjectured.  Another surprising result is that the most Wigner-negative Dicke state is not always the equatorial $\ket{j,m=0,\pm 1/2}$ state, and that this stems from a nontrivial dependence on spin itself.

It is interesting to compare Wigner negativity and entanglement as measures of nonclassicality.  It is known that within a Dicke basis of arbitrary spin $j$, the equatorial states are maximally entangled as seen by the entanglement entropy across arbitrary bipartitions of $2j$ qubits \cite{moreno_all_2018}.  The geometric measure of entanglement similarly witnesses the equatorial Dicke state as having the most entanglement within the Dicke basis for arbitrary $j$ \cite{aulbach_maximally_2010}.  These are in contrast with Wigner negativity, where the most Wigner-negative Dicke state bifurcates from $\ket{j,0}$ to $\ket{j,\pm \, m'_j}$ around $j\gtrsim 30$.  On the other hand, the entanglement entropy of sequences $\{\ket{j,j-n}\}_j$ across specifically half-bipartitions (i.e.\ an even $j:j$ splitting of $2j$ qubits for integer spin) seem to approach a constant value for any fixed $n$ \footnote{It should be mentioned however that the entanglement entropy of the sequence $\{\ket{j,j-n}\}_j$ across a $1:(2j-1)$ qubit bipartition vanishes in the limit of infinite $j$ for any fixed $n$ \cite{moreno_all_2018}, and thus does not approach a finite constant nonzero value like the half-bipartition entropy.} \cite{stockton_characterizing_2003}.  In a similar fashion as established here, the Wigner negativity of $\{\ket{j,j-n}\}_j$ also approaches a constant value; i.e. that of the number state $\ket{n}$ for any fixed $n$.

Thus the behavior of spherical Wigner negativity qualitatively agrees with entanglement entropy when considering half-bipartitions of sequences $\{\ket{j,j-n}\}_j$, but disagrees on which Dicke basis element is the most nonclassical.  Furthermore, the GHZ state is relatively less Wigner negative than most Dicke states with the same spin value as seen in Fig.\ \ref{fig:dicke-and-fock-negativities}. This suggests that Wigner negativity and entanglement capture different aspects of the nonclassicality of states.

The nontrivial and spin-dependent behaviour of SU(2)-covariant Wigner negativity suggests a complex structure to the nonclassicality of symmetric qubit ensembles as their system size grows.  This highlights the influence that geometric properties of classical phase space, and so more importantly the associated dynamical symmetry, has on the nonclassicality of quantum states living on it.

\section*{Acknowledgements}
JD acknowledges the support of the Natural Sciences and Engineering Research Council of Canada (NSERC).
This research was supported in part by Perimeter Institute for Theoretical Physics. Research at Perimeter Institute is supported in part by the Government of Canada through the Department of Innovation, Science and Economic Development Canada and by the Province of Ontario through the Ministry of Colleges and Universities.  

\appendix
\section{Lower bound on spin cat states}\label{sec:wigner-analytics}
Here, we briefly expand on the steps taken to obtain the lower bound on the GHZ/N00N state negativity.  We presented the SU(2)-kernel as derived by Heiss and Weigert \cite{heiss_discrete_2000} in Eq.\ \eqref{eq:su2-kernel-diagonal} due to their emphasis on the rotation-independence of the eigenvalues along any quantization axis $\bm{n}$. However, for calculations along specifically the north-south axis, we used an equivalent form of the kernel derived in \cite{varilly_moyal_1989} and presented in \cite{klimov_generalized_2017}:
\begin{equation}
\label{eq:Klimov-kernel}
    \Delta(\Omega) = \sqrt{\frac{4\pi}{2j+1}} \sum_{l=0}^{2j} \sum_{k=-l}^l Y^*_{lk}(\Omega) T^{(j)}_{lk}
\end{equation}
where
\begin{equation}
    T^{(j)}_{lk} = \sqrt{\frac{2l+1}{2j+1}}\sum_{n,n'=-j}^j \cg{j}{n}{l}{k}{j}{n'} \ketbra{j,n'}{j,n}
\end{equation}
are spherical tensor operators and $Y^*_{jm}(\Omega)$ is the complex conjugate of the standard spherical harmonics.  One may readily verify that a Dicke state expectation of this kernel \eqref{eq:Klimov-kernel} reduces to the Dicke state Wigner function of Eq.\ \eqref{eq:dicke-state-wigner-function}, with the $m=j$ case corresponding to the north pole spin coherent state representative discussed in section \eqref{SEC:scs}.  

The Wigner function of the spin cat state
\begin{equation}
\ket{\psi} = \cos\bp{\frac{\vartheta}{2}} \ket{j,j} + e^{i\varphi}\sin\bp{\frac{\vartheta}{2}}\ket{j,-j}
\end{equation}
is split into two antipodal spin coherent state contributions and two cross-term contributions $\langle j,\pm j | \Delta (\Omega) | j, \mp j \rangle$, with the latter pair containing the characteristic interference pattern.  The first cross-term evaluates to
\begin{align}
&\langle j, j | \Delta (\Omega) | j, - j \rangle = \nonumber \\
&\quad \sqrt{\frac{4\pi}{2j+1}} \sum_{l=0}^{2j} \sum_{k=-l}^l Y^*_{lk}(\Omega) \bs{\sqrt{\frac{2l+1}{2j+1}} \cg{j}{-j}{l}{k}{j}{j} \delta_{k,2j}}
\end{align}
which is nontrivial for simultaneous $l = 2j$ and $k = 2j$, giving
\begin{equation}
\langle j, j | \Delta (\Omega) | j, - j \rangle = \sqrt{\frac{4\pi}{2j+1}} (-1)^{2j} Y^*_{2j,2j}
\end{equation}
because $\cg{j}{-j}{2j}{2j}{j}{j} = (-1)^{2j}\sqrt{\frac{2j+1}{4j+1}}$.  The entire interference contribution becomes
\begin{align}
& \quad \sin \vartheta \sqrt{\frac{2j+1}{4j+1}} \Re\bs{e^{i\varphi} (-1)^{2j} Y^*_{2j,2j}(\Omega) } \\
&= \sin \vartheta N_j \sin^{2j}(\theta) \cos(2j\phi - \varphi) \label{eq:appendix-ghz-interference}
\end{align}
where $N_j = \frac{1}{2^{2j} (2j)!} \sqrt{(4j+1)!/(2j+1)}$ and we have used the relation $Y_{l,l}(\theta,\phi) = \frac{(-1)^{l}}{2^l l!} \sqrt{\frac{(2l+1)!}{4\pi}}\sin^l(\theta)e^{ il\phi}$.  We note that an asymptotic form of the GHZ state, valid only for integer spin, has been given in \cite{klimov_su2_2002}.

The lower bound on the Wigner negativity is obtained by ignoring the interaction between the equatorial fringes and the antipodal spin coherent contributions.  Restricting to the interference term \eqref{eq:appendix-ghz-interference}, we see that the nonnegative sine function over the polar angle $\theta\in[0,\pi]$ is symmetric about the equator, while the cosine function periodically splits the azimuthal dependence into $2j$ identical regions (we also set $\varphi=0$).  Focusing on where the cosine becomes negative, we integrate \eqref{eq:appendix-ghz-interference} over the region $[0,\frac{\pi}{2}] \cup \frac{1}{2j}\bs{\frac{\pi}{2},\frac{3\pi}{2}}$, and then multiply the result by $2\cdot2j = 4j$:
\begin{equation}
    4j \sin \vartheta N_j \frac{2j+1}{4\pi} \int_0^{\pi/2} \sin^{2j}(\theta)\sin\theta \de\theta \int_{\pi/4j}^{3\pi/4j} \cos(2j\phi) \de\phi.
\end{equation}
The azimuthal component integrates to $-1/j$.  The polar component must employ the recursive relation
\begin{equation}
    \int_0^{\pi/2} \sin^{n}(\theta) \de \theta = \frac{n-1}{n}\int_0^{\pi/2} \sin^{n-2}(\theta)\de \theta
\end{equation}
for integer $n$, which leads to the known identity
\begin{equation}
    \int_0^{\pi/2} \sin^n(\theta)\de\theta = \tilde{\sigma}(n) \frac{(n-1)!!}{n!!}
\end{equation}
where $\tilde{\sigma}(n)=\pi/2$ for even $n$ and $\tilde{\sigma}(n)=1$ for odd $n$.  The upper bound then becomes
\begin{align}
    \sin \vartheta N_j \frac{\sigma(j)}{\pi} \frac{(2j)!!}{(2j-1)!!}
\end{align}
where $\sigma(j) = 1$ for integer spin, $\sigma(j) = \pi/2$ for half-integer spin, and the result has been multiplied by $-1$ to yield a positive number.  The explicit special cases are obtained from the identities $n!! = 2^k k!$ for even integer $n=2k$ and $n!! = (2k)!/(2^k k!)$ for odd integer $n=2k-1$.

%\bibliography{main-library2}


\end{document}